\documentclass[12pt]{iopart}

\usepackage{graphicx}
\begin{document}

\title[]{How accurately can the Extended Thomas-Fermi method describe the inner crust of a neutron star?}

\author{M. Shelley and A. Pastore}

\address{ Department of Physics, University of York, Heslington, York, Y010 5DD, United Kingdom}
\ead{mges501@york.ac.uk}
\vspace{10pt}
\begin{indented}
\item[]\today
\end{indented}

\begin{abstract}
We perform a systematic comparison between the results obtained by solving fully self-consistently the Hartree-Fock-Bogoliubov equations, and those obtained using the semi-classical Extended Thomas-Fermi method, for various Wigner-Seitz cells within the inner crust of a neutron star. 
The lack of pairing correlations in the semi-classical approach leads to a large discrepancy between the two approaches. This discrepancy is well beyond the error of the quantum-mechanical calculation, and is related to spurious shell effects in the neutron gas.
\end{abstract}

%
%
%
%
%

\section{Introduction}

The recent detection of two neutron star (NS) mergers was made using the innovative measurements of gravitational~\cite{abbott2017gw170817} and electromagnetic~\cite{cowperthwaite2017electromagnetic} waves. The combination of the two techniques opens up a new era for observing these interesting astrophysical objects, and provides a new set of tools to better understand the physical properties of a NS~\cite{cha08,haensel2007neutron}.

The key ingredient to describe the physics of a NS is the Equation of State (EOS), \emph{i.e.} the relation between the pressure and the matter density~\cite{ste10}. Given the strong pressure gradient, the matter within the star is arranged in layers with different characteristics. Going from the outside (low-density) to the centre of the star (high density), we find two main regions: the crust and the core.
The matter in the crust consists of neutron rich nuclei surrounded by a free electron gas (outer crust) and by a free neutron gas (inner crust). At baryonic densities of $\rho_B\approx0.09$ fm$^{-3}$~\cite{gonzalez2017higher,xu2009locating,chamel2015brussels}, nucleons no longer form a cluster, but behave as a Fermi liquid. This region of the star is the core; this extends over a large density range, reaching values that are typically 3 to 4 times the standard saturation density found within nuclei.
Consequently, the composition of the core is not known in detail, and several models have been suggested~\cite{zho04,heb10,dou11,ste05,for14,sha15,vidana2018d,chatterjee2016hyperons}.

To build an \emph{universal} EOS, all the different layers of the NS should be described coherently, without matching different models for the different regions of the star. This \emph{matching} problem has been discussed in~\cite{fortin2016neutron}, and leads to additional uncertainties in predictions of the NS maximum radius. 

To avoid such a shortcoming, several groups have investigated a \emph{unified} EOS, \emph{i.e.} using the same model to describe all layers of the star~\cite{douchin2001unified,sha15,cha11,kumar2018new}.
The major difficulty in achieving this goal is obtaining a correct description of the inner crust region~\cite{cha08}.
As illustrated in the pioneering work of Negele and Vautherin~\cite{negele1973neutron}, this region of the crust is composed of very neutron-rich nuclei, arranged in a crystalline structure and surrounded by a gas of superfluid neutrons and ultrarelativistc electrons~\cite{pastore2011superfluid}.

The tool of choice to describe this region is Nuclear Energy Density Functional theory~\cite{bender2003self}. By solving the Hartree-Fock-Bogoliubov (HFB) equations within a Wigner-Seitz (WS) cell~\cite{chamel2007validity}, one can obtain the detailed structure of this region as a function of the baryonic density~\cite{pastore2017new,baldo2005role,grill2011cluster}.
This procedure may be very time consuming and numerically inaccurate~\cite{baldo2006role}, due to the particular choice of how to treat the neutron gas states. To avoid this issue, several groups have opted for a simpler treatment of the system using the semi-classical Thomas-Fermi (TF) approximation~\cite{ring2004nuclear}.
In the present article, we perform a systematic comparison of the extended TF and Hartree-Fock-Bogoliubov methods for the inner crust of a NS. Such a comparison has been routinely performed for finite nuclei, but never for the inner crust of a NS while controlling for the different aspects of the calculations.

The article is organised as follows: in Section\ref{Sec:HFB}, we present the HFB equations, while in Sec.\ref{sec:TF} we introduce the ETF approximation. In Sec.\ref{sec:ic}, we illustrate our findings and finally we provide our conclusions in Sec.\ref{sec:conc}.

\section{Hartree-Fock-Bogoliubov}\label{Sec:HFB}

A simple way to describe the properties of a nucleus embedded in a neutron gas is to solve the  Hartree-Fock-Bogoliubov equations together with an effective interaction. They read~\cite{ring2004nuclear}

\begin{eqnarray}\label{HFBeq}
  \sum_{n'}(h_{n'nlj}^q- \varepsilon_{F,q})U^{i,q}_{n'lj}+\sum_{n'}\Delta_{nn'lj}^qV^{i,q}_{n'lj}=E^{q}_{ilj}U^{i,q}_{nlj}\;, & &
     \nonumber \\
  \sum_{n'}\Delta^q_{nn'lj}U^{i,q}_{n'lj} -\sum_{n'}(h^q_{n'nlj}- \varepsilon_{F,q})V^{i,q}_{n'lj}  =E^{q}_{ilj}V^{i,q}_{nlj}\;. & &
\end{eqnarray}

$\varepsilon_{F,q}$ is the Fermi energy and $q$ stands for neutrons (n) and protons (p).
In the present work, we assume that the system is spherically symmetric, so we used the standard notation $nlj$ for the single-particle states with radial quantum number $n$, orbital angular momentum $l$ and
total angular momentum $j$.
$U^{i,q}_{nlj}$ and $V^{i,q}_{nlj}$ are the Bogoliubov amplitudes for the $i$-th quasiparticle
of energy $E^{q}_{ilj}$.
We refer to Refs~\cite{pastore2011superfluid,pastore2012superfluid} for a detailed discussion on the adopted numerical techniques used to solve these equations. An important aspect of our method is that we discretise the continuum states, by setting Dirichlet-Neumann mixed boundary conditions at the edge of the WS cell. There are two cases to consider. The first is where we impose that even-parity wave functions vanish at edge of the box $R_{B}$ and that the first derivatives of odd-parity wave functions vanish at $R_{B}$; we call this Boundary Conditions Even (BCE). The second case is where the two parity states are treated in the opposite way, which we call Boundary Conditions Odd (BCO).
Other boundary conditions have been used in the literature to properly treat continuum states~\cite{chamel2007validity,grasso2001pairing,schuetrumpf2015twist}.
For simplicity, we limit ourselves to the most common implementation of boundary conditions, which have been used previously to perform systematic calculations of inner crust properties~\cite{grill2011cluster,baldo2006role,than2011wigner,sandulescu2004nuclear} 
In the limit of vanishing pairing, Eqs.~\ref{HFBeq} reduce to the simple Hartree-Fock (HF) case.

For the particle-hole channel, we use the SLy4 functional~\cite{chabanat1998skyrme}, while for the pairing sector we adopt a simple density dependent delta interaction

\begin{eqnarray}\label{vpair}
v_{pair}(\mathbf{r}_1,\mathbf{r}_2)=V_0\left[1-\eta \left(\frac{\rho_B\left(\frac{\mathbf{r}_1+\mathbf{r}_2}{2}\right)}{\rho_0} \right)^\alpha \right]\delta(\mathbf{r}_1-\mathbf{r}_2);,
\end{eqnarray}

\noindent The interaction strength is fixed to $V_0=-430$ MeVfm$^3$; while the other parameters take the values $\eta=0.7, \alpha=0.45$ and $\rho_0=0.16$ fm$^{-3}$. To avoid the ultraviolet divergence related to the zero-range nature of the pairing interaction~\cite{bulgac2002renormalization}, we adopt a cut-off of 60 MeV in the quasi-particle spectrum. See Ref.~\cite{pastore2011superfluid} for more details.

\section{Extended Thomas Fermi}\label{sec:TF}

Within the Skyrme model~\cite{skyrme1956cvii}, it is possible to write the total energy as an energy density functional

\begin{equation}\label{eq:functional}
E=\int \mathcal{E}(\rho_q(\mathbf{r}),\tau_q(\mathbf{r}),\vec{J}_q(\mathbf{r}))d^3\mathbf{r}\;,
\end{equation}

\noindent which depends on the local matter densities $\rho_q(\mathbf{r})$, the kinetic energy densities $\tau_q(\mathbf{r})$, and the spin current densities $\vec{J}_q(\mathbf{r})$. Other densities may also occur, but in the present article we consider only the time-even sector~\cite{lesinski2007tensor} of a \emph{standard} Skyrme functional~\cite{becker2017solution,carlsson2008local}. 

Within the HFB scheme, the densities are calculated using the quasi-particle wave functions~\cite{lesinski2006isovector}, while in the semi-classical approach they are parameterised with modified Fermi-Dirac functions 

\begin{eqnarray}\label{TF:dens}
\rho_q(r)=\frac{\rho_0^q}{\left[1+\exp\left( \frac{r-R_0^q}{a_q}\right) \right]^{\gamma_q}}+\rho^q_{gas}\;.
\end{eqnarray}

\noindent The parameters $R_0^q,a_q,\rho_0^q,\gamma_q,\rho^q_{gas}$ are fitted to reproduce the quantal densities, under the constraint of keeping the correct numbers of neutrons and protons.
The parameter $\rho^q_{gas}$ is added to the standard form~\cite{bartel2002nuclear} to account for the presence of a neutron Fermi gas in the cell.
Using the Wigner-Kirkwood (WK) expansion~\cite{ring2004nuclear}, one gets expressions for the kinetic and spin current densities.
For brevity, we do not give the expressions here, but we refer to Ref.~\cite{bartel2002nuclear} where all equations are explicitly written in great detail.
The WK expansion can be truncated at a given order; here we have decided to consider terms only up to second order. The terms from this expansion lead to corrections to the kinetic and spin current densities. To distinguish from a simple Thomas Fermi approach, it takes the name of \emph{extended Thomas Fermi} (ETF)~\cite{onsi2008semi,martin2015liquid} in the literature.

\begin{figure}[h]
    \begin{center}
    \includegraphics[width=0.7\textwidth]{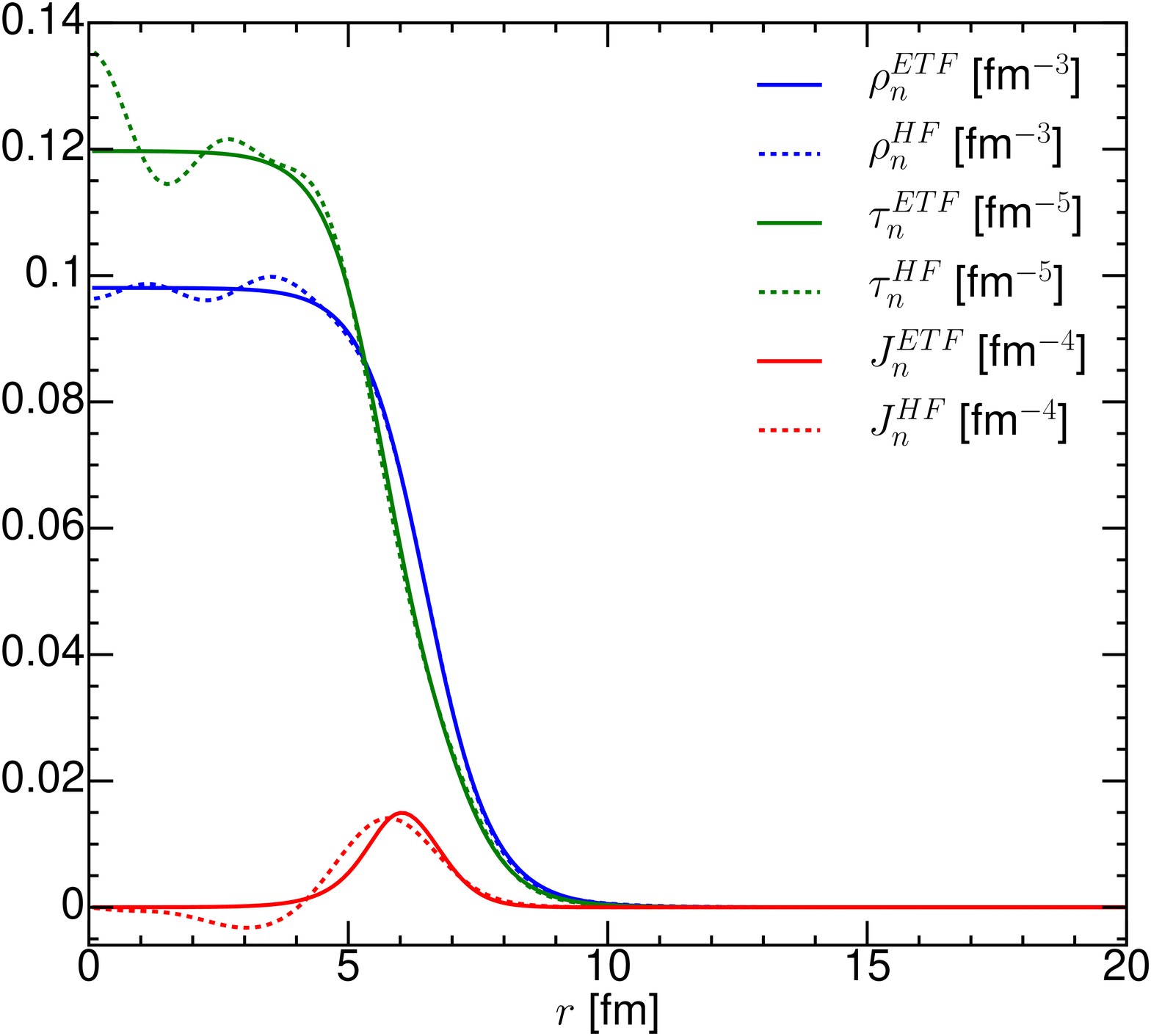} 
        \caption{(Colors online) Neutron densities obtained using a full HF calculation (dashed) and the ETF method (solid) for $^{176}$Sn. See text for details.}
            \label{Dens176Sn}
    \end{center}
\end{figure}

\begin{figure*}[h]
   \begin{center}
        \includegraphics[width=0.49\textwidth]{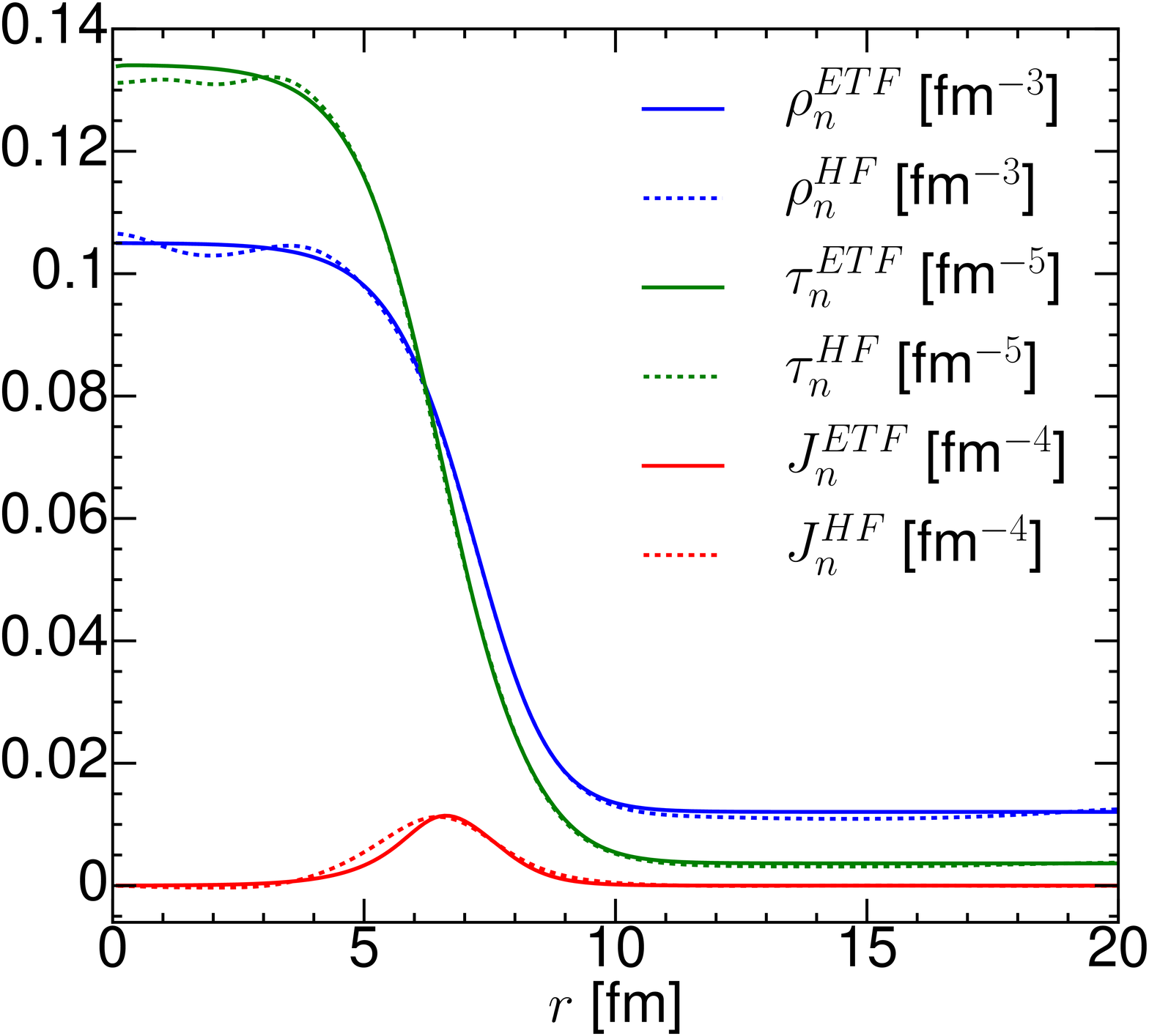}
        \includegraphics[width=0.49\textwidth]{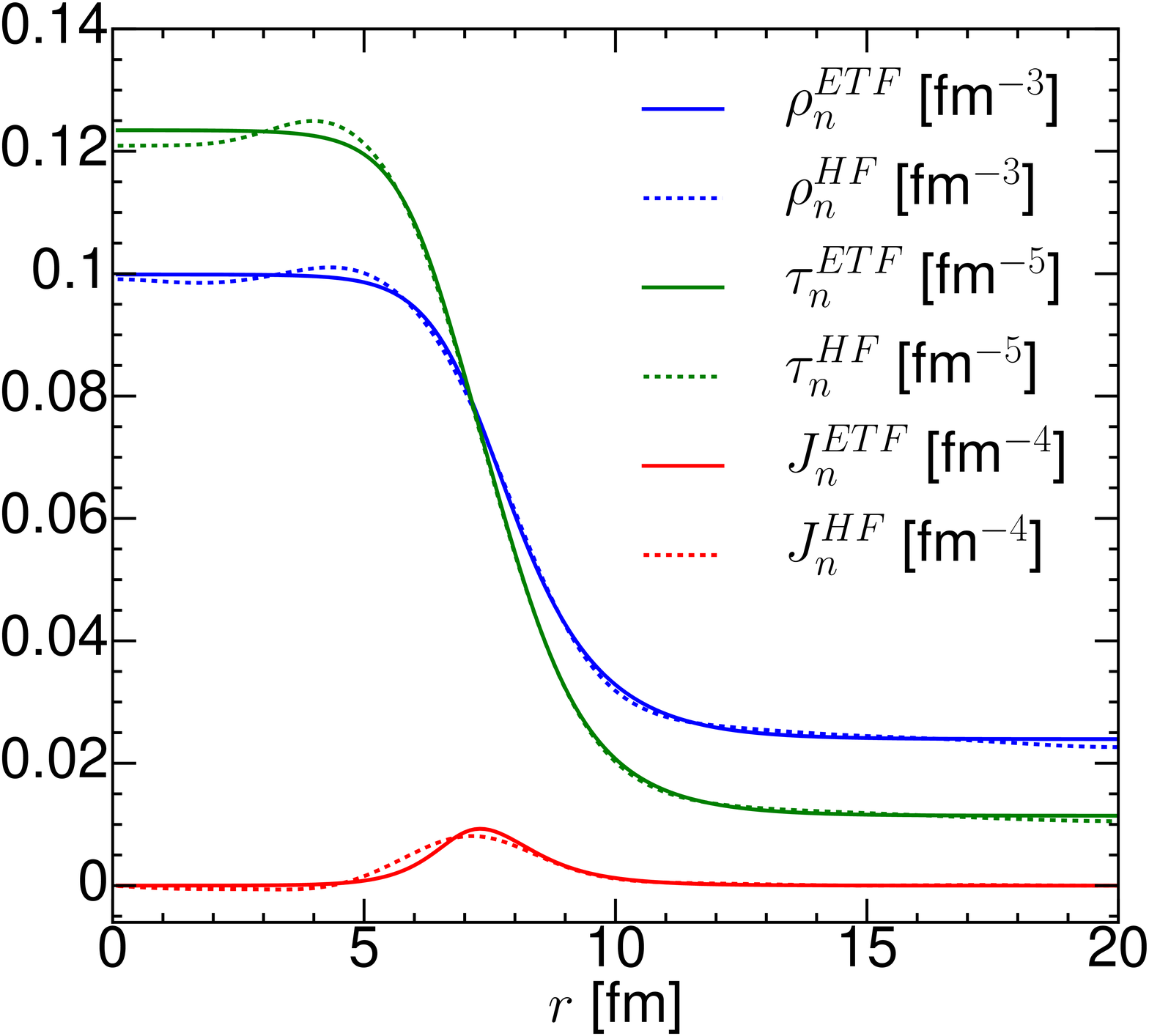} 
            \caption{(Colors online) Neutron densities obtained using full HF calculations (dashed) and using the ETF method (solid), for the WS cell with Z=50, and $\rho_b=0.012$ fm$^{-3}$ (left panel) and $\rho_b=0.024$ fm$^{-3}$ (right panel). See text for details.}
                \label{DensXX}
    \end{center}
\end{figure*}

\noindent In Fig.~\ref{Dens176Sn}, we compare the HF and semi-classical neutron densities of $^{176}$Sn.
The parameters of Eq.~\ref{TF:dens} are fitted on the HF density $\rho_q^{HF}$, while the kinetic $\tau^{ETF}_q$ and spin-current $J_q^{ETF}$ densities have been derived using the ETF method.
We observe that the ETF method reproduces the main features of the HF results very well, except for the oscillating behavior in the interior, which is related to the underlying shell structure.
In Fig.~\ref{DensXX}, we perform the same comparison, but now for two fictitious WS cells of size $R_{B}=60$ fm, with $Z=50$ protons, and baryonic densities of $\rho_B=0.012$ fm$^{-3}$ and $\rho_B=0.024$ fm$^{-3}$ respectively.
These two values roughly correspond to the middle of the crust and to the end point just before the transition into uniform matter.
We observe that, for these denser systems, the ETF method is approximating the densities even better, compared with the results provided in Fig.~\ref{Dens176Sn}.

The higher-order terms in the ETF method comprise linear combinations of derivatives of the density given in Eq.~\ref{TF:dens} and of the effective mass. In Ref.~\cite{bartel2002nuclear}, it was shown that they improve very little the agreement between the HF and semi-classical densities in finite nuclei. For the sake of simplicity we exclude them in the present work.

\section{Inner crust}\label{sec:ic}

\begin{figure*}[h]
    \centering
        \includegraphics[angle=0,width=0.7\textwidth]{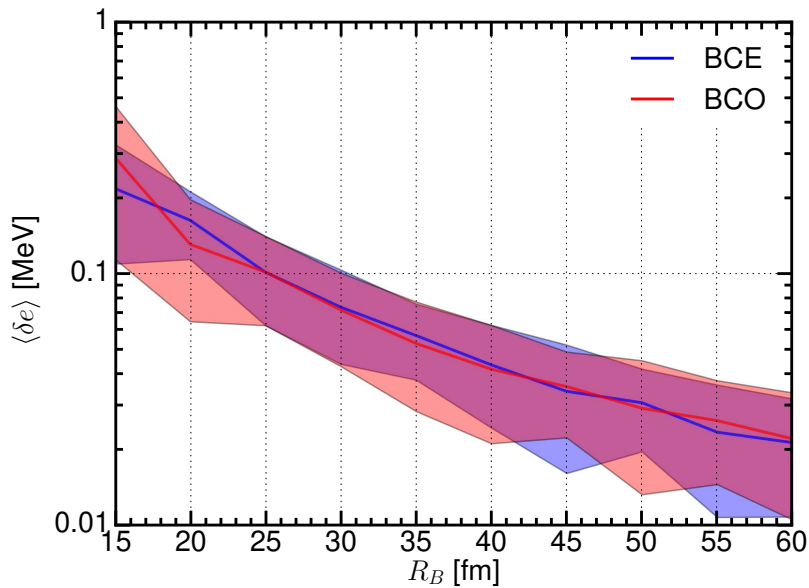}
            \caption{(Colors online)  Error bar as a function of the size of the box used to perform HF calculations and for two sets of boundary conditions. See text for details.}
                \label{FiniteSize}
\end{figure*}

To benchmark the accuracy of the two methods, we first revise the quality of the present HFB calculations. Following the procedure illustrated in Refs.~\cite{pastore2017new,margueron2008equation}, we calculate the energy per particle of a system of neutrons without pairing at a given density, by solving Eqs.~\ref{HFBeq} in a spherical box of various sizes.
In the limit of a very large box radius, these results should reproduce the analytical values of the energy per neutron $\frac{E_{PNM}}{N}$ obtained by solving Hartree-Fock equations in infinite pure neutron matter (PNM).
In Fig.~\ref{FiniteSize}, we show the evolution of the energy difference

\begin{equation}
    \delta e=\left|\frac{E_{HF}}{N}-\frac{E_{PNM}}{N} \right|\;,
\end{equation}

\noindent as a function of the size of the box $R_B$. We have performed two sets of calculations for the two sets of boundary conditions.
As shown in Ref.~\cite{pastore2017new}, there is a weak density dependence on $\delta e$ for a fixed value of $R_B$.
Since we are not interested in estimating a very precise error, we take the averages, not showing explicitly this density dependence, to illustrate the error $\langle \delta e \rangle$ as a function of the box size in the form of bands.
The solid line in the middle of each band is to guide the eye and represents an average value.
From Fig.~\ref{FiniteSize}, we can read off the error related to the continuum discretisation carried out in the HF calculations. For $R_B=60$ fm we have $\langle \delta e\rangle\approx20$ keV.

Having estimated the HF error, we can proceed to a more detailed comparison between ETF and HF(B).
To this purpose, we run a series of HF calculations (\emph{i.e.} with no pairing correlations), from the drip line nucleus $^{176}$Sn up to the limit of existence for the crust, keeping the number of protons fixed.
These WS cells all have the same size $R_B=60$ fm and proton number $Z=50$, as in Ref~\cite{grasso2008low}.
At present, we are not seeking a realistic description of the inner crust, so we need not consider these cells at $\beta$ equilibrium~\cite{negele1973neutron}.
For each WS cell, we also carry out an ETF calculation using the HF densities as an input.
We calculate the total energy of the system using Eq.~\ref{eq:functional}.
To take into account the physical shell effects present in the cluster, we use the Strutinsky integral correction for protons~\cite{onsi2008semi,pea12}.
We are interested in the discrepancy between the two methods: how different is the calculated energy per particle from the ETF and HF methods, for the relevant baryonic densities?

\begin{figure*}[h]
    \centering
        \includegraphics[angle=0,width=0.7\textwidth]{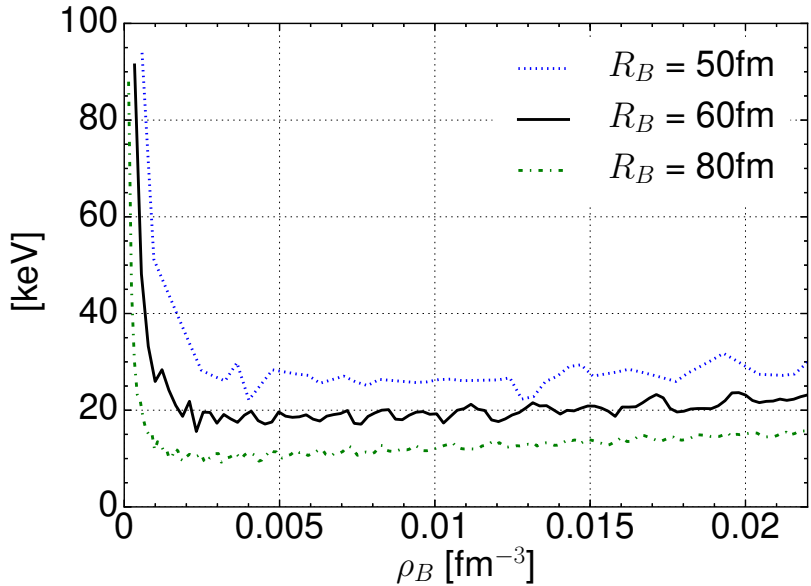}
            \caption{(Colors online) Energy per particle difference as a function of the baryonic density of the system for the EFT-HF case and three different box sizes: $R_B=50$ fm (dotted line), $R_B=60$ fm (solid line) and $R_B=80$ fm (dash-dotted line). See text for details.}
                \label{Err:ETF-HFB_BOXEVO}
\end{figure*}

In Fig.~\ref{Err:ETF-HFB_BOXEVO}, we show the evolution of the energy per particle difference between the ETF and HF methods, for different box sizes.
We observe that, for very low-density WS cells, the discrepancy between the two methods can be as high as $\approx$ 100 keV per particle.
Above densities of $\rho_B\approx0.002$ fm$^{-3}$ it drops quickly and approaches a near-constant value at high densities.
However, this constant discrepancy at large densities is dependent on the box size: for $R_B=50$ fm the difference is $\approx$ 30 keV, for $R_B=60$ fm it is $\approx$ 20keV, and for a very large box of $R_B=80$ fm the error falls to $\approx$ 15 keV.
These values are compatible with the errors of the HF method as extracted from Fig.~\ref{FiniteSize}, showing that the two methods are in good agreement (within their error bars) at these densities.
Importantly, this dependency of the discrepancy on the box size demonstrates that the discrepancy arises from a poor treatment of the neutron gas states in the HF method.

\begin{figure*}[h]
    \centering
    \includegraphics[angle=0,width=0.7\textwidth]{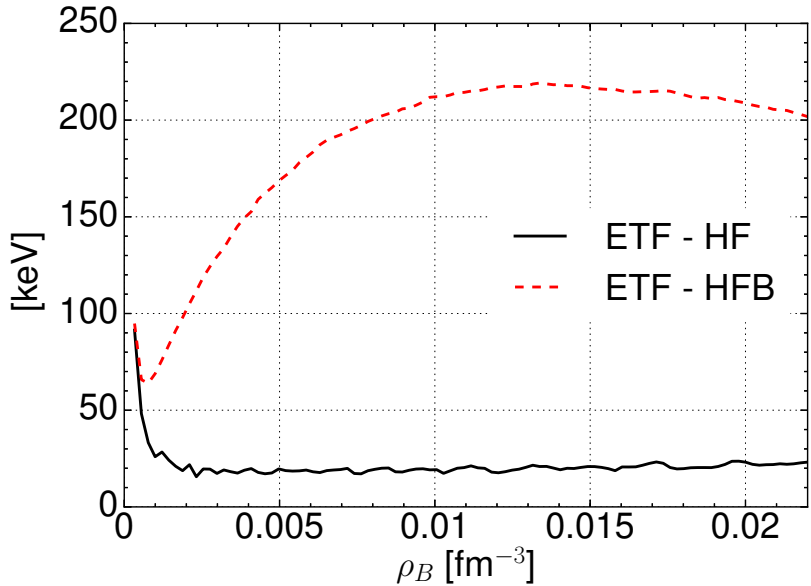}
        \caption{(Colors online) Energy per particle difference, as a function of the baryonic density of the system, for the EFT-HF case (solid line) and ETF-HFB (dashed). See text for details.}
            \label{Err:ETF-HFB}
\end{figure*}

In Fig.~\ref{Err:ETF-HFB}, we show again the difference between the energy per particle obtained with the ETF method as detailed before and that obtained with the fully self-consistent HF calculation, as a function of the baryonic density of the system (solid black line, labelled `ETF-HF', the same as the one labelled `$R_B=60$ fm' in Fig.~\ref{Err:ETF-HFB_BOXEVO}).
On the same figure, we now show the difference between the ETF energy per particle and that obtained with a full HFB calculation using the pairing interaction defined in Eq.~\ref{vpair}.
In this case the energy difference follows a different trend: it starts decreasing at very low density, but when pairing switches on it starts increasing again, reaching a maximum of $\approx$220 keV per particle.
At higher densities, when the pairing starts decreasing, the energy difference starts decreasing again.
Except at very low densities, the discrepancy between the two methods is one order of magnitude worse than in the non-superfluid case.
This clearly shows that neglecting pairing correlations for neutrons with ETF leads to a much larger error than the one related to discretisation effects and the one based on how we treat continuum states in the neutron gas with HFB.

\section{Conclusions}\label{sec:conc}

In this article, we have presented our application of the Extended Thomas Fermi method. The ETF method is a valuable tool to perform systematic calculations of properties of WS cells within the inner crust of a NS.

We have built a series of WS cells with a fixed number of protons $Z=50$ and performed a systematic comparison between the Hartree-Fock and ETF results. We have seen that at very low densities ($\rho_B\le0.002$ fm$^{-3}$) ETF has a remarkably large error of the order $\approx100-150$ keV per particle.
At larger densities the discrepancy between the two calculations decreases and becomes compatible with the estimated error bar on Hartree-Fock calculations.

We have also compared the ETF results with full Hartree-Fock-Bogoliubov calculations: most ETF calculations simply neglect the pairing energy contribution from the neutrons. We have shown here that such an approximation leads to systematic error of the order of $\approx200$ keV per particle. This is one order of magnitude larger than the standard HFB error, which comes from the artificial discretisation of continuum states due to particular choice of boundary conditions~\cite{pastore2017new,pastore2017impact}.

From our analysis we thus conclude that the ETF method can be considered as a valuable tool only if pairing correlations are also included for neutrons~\cite{schuck2013thomas,pearson2015role}. Given previous analysis of the errors incurred from pairing approximations in a WS cell (see Fig.~2 in Ref.~\cite{pastore2017impact}), we expect that the difference in energy per particle to drop to $\approx20-30$ keV per particle.
By combining the low computational cost of ETF calculations with modern statistical technique of Gaussian Process Emulation (GPE)~\cite{shelley2018advanced}, we plan to perform a full analysis of the equation of state of the NS inner crust, including temperature effects.

\ack

This work was supported by the STFC grants ST/M006433/1
and ST/P003885/1.

\section*{References}
\bibliographystyle{iopart-num}

\bibliography{biblio}

\end{document}